\documentclass[usenatbib]{mn2e}
\usepackage{graphicx}
\usepackage{ifthen}

\def\ltsima{$\; \buildrel < \over \sim \;$}
\def\simlt{\lower.5ex\hbox{\ltsima}}
\def\gtsima{$\; \buildrel > \over \sim \;$}
\def\simgt{\lower.5ex\hbox{\gtsima}}
\def\kms{{\rm\,km\,s^{-1}}}

\def\masyr{{\rm\,mas/yr}}
\def\kpc{{\rm\,kpc}}

\def\msun{{\rm\,M_\odot}}

\def\AA{$\; \buildrel \circ \over {\rm A}$}

\def\UseFigs{1}

\def\deg{^\circ}

\def\s{\ifmmode \widetilde \else \~\fi}
\def\={\overline}

\def\spose#1{\hbox to 0pt{#1\hss}}

\def\lta{\mathrel{\spose{\lower 3pt\hbox{$\mathchar"218$}}
     \raise 2.0pt\hbox{$\mathchar"13C$}}}
\def\gta{\mathrel{\spose{\lower 3pt\hbox{$\mathchar"218$}}
     \raise 2.0pt\hbox{$\mathchar"13E$}}}
\def\Dt{\spose{\raise 1.5ex\hbox{\hskip3pt$\mathchar"201$}}}    
\def\dt{\spose{\raise 1.0ex\hbox{\hskip2pt$\mathchar"201$}}}    

\def\dotsfill{\leaders\hbox to 1em{\hss.\hss}\hfill}

\loadboldmathitalic 
\title[Kinematics of the Canis Major dwarf galaxy] {A radial velocity survey of low Galactic latitude structures: I. Kinematics of the Canis Major dwarf galaxy}
\author[N. F. Martin et al.] {N. F. Martin$^{1}$, R. A. Ibata$^{1}$, B. C. Conn$^{2}$, G. F. Lewis$^{2}$ , M.
Bellazzini$^{3}$ \&
\newauthor M. J. Irwin$^{4}$\\
$^{1}$ Observatoire de Strasbourg, 11, rue de l'Universit\'e, F-67000, Strasbourg, France\\
$^{2}$ Institute of Astronomy, School of Physics, A29, University of Sydney, NSW 2006, Australia\\
$^{3}$ INAF - Osservatorio Astronomico di Bologna, Via Ranzani 1, 40127, Bologna, Italy\\
$^{4}$ Institute of Astronomy, Madingley Road, Cambridge, CB3 0HA, U.K.\\
}

\date{\today}
\begin{document} 
\maketitle 
\begin{abstract} 
As part of a radial velocity survey of low Galactic latitude structures that we undertook with the 2dF spectrograph on
the AAT, we present the radial velocities of more than 1500 Red Giant Branch and Red Clump stars towards the centre of
the Canis Major dwarf galaxy. With a mean velocity of $72\pm7\kms$ at a Heliocentric distance of $5.5\kpc$ and
$114\pm2\kms$ at $8.5\kpc$, these stars present a peculiar distance -- radial velocity relation that is unlike that
expected from thin or thick disc stars. Moreover, they belong to a kinematically cold population with an intrinsic
dispersion that may be as low as $11_{-1}^{+3}\kms$. A comparison of the velocity distribution obtained in this work
with previous studies, shows the importance of using our new reduction pipeline and averaging the velocities obtained
from different templates.\\
The radial velocity distribution is used to select Canis Major stars in the UCAC2.0 proper
motion catalogue and derive proper motions in Galactic coordinates of $(\mu_l,\mu_b)= (-3.6\pm0.8\masyr,
1.5\pm0.4\masyr)$ for the dwarf galaxy, which after correcting for the reflex solar motion along this line-of-sight
gives $(\mu_l',\mu_b')= (-6.8\pm0.8\masyr, 0.8\pm0.4\masyr)$, corresponding to a prograde orbit with a tangential
velocity of $\sim235\kms$ at the average distance of $\sim7.2\kpc$. All these kinematic constraints can be reproduced in
simulations of the accretion of a dwarf onto the Galactic disc. Such a process could also be responsible for the
Monoceros Ring that has recently been shown to encompass the Galactic disc. However, without constraints on the
kinematics of the tidal arms emerging from the Canis Major dwarf, it is not yet possible to definitively prove a link
between the two structures.

\end{abstract}


\section{Introduction}

The advent of all sky surveys is revealing numerous structures toward the edge of the Galactic disc. The
SDSS revealed a ring-like structure, the so-called Monoceros ring (Mon ring), that was later shown to encompass part of
the Galactic disc \citep[e.g.][]{newberg02,yanny03,ibata03,crane03,conn05}. Similarly, the 2MASS infrared catalogue was
used to unveil the existence of a diffuse structure in the direction of the Triangulum and Andromeda constellations
\citep{rocha-pinto04} and to reveal the presence of a dwarf galaxy below the disc in the CMa constellation
\citep[hereafter Paper~I]{martin04a}.

This latter structure was first identified as an overdensity of Red Giant Branch (RGB) stars located just below the
Galactic disc at $(l,b)\sim(240\deg,-8\deg)$. The analysis of archival
Colour-Magnitude Diagrams (CMD) of open clusters that are fortuitously located around this region confirmed this stellar
overdensity extends over $20\deg$ in Galactic longitude \citep[hereafter Paper~II]{bellazzini04} while main sequence
fitting revealed this population has an intermediate age (4-10 Gyr). Applying a Tip of the Red Giant Branch algorithm
yielded a distance of $7.2\pm0.3\kpc$ for the dwarf \citep[hereafter Paper~III]{martin04b}, making it the closest
galaxy from the Sun.  Interestingly, Canis Major lies at a comparable Galactocentric distance to the Sagittarius dwarf.

As an alternative interpretation, \citet{momany04} explain this stellar structure as a signature of the Galactic warp.
Using the UCAC2.0 proper motion catalogue, they derive a proper motion that, within sizable uncertainties, is compatible
with a Galactic population on a prograde disc-like orbit:
$(\mu_{\alpha}\cos(\delta),\mu_{\delta})=(-1.7\pm2\masyr,3.1\pm2\masyr)$. However, comparison of deep optical
photometry CMDs with Galactic models reveals the structure is incompatible with known Galactic structure (Paper II) and
yields a narrow extent on the line of sight, with a FWHM of only $1.92\kpc$ \citep{martinez-delgado05}, difficult to
achieve with a conventional warp. Moreover, the radial velocities of a small sample of RGB stars in CMa show they
belong to a kinematically cold population that, if on a circular orbit, would rotate around the Milky Way at a
rotational velocity of $\sim160\kms$ that is low for disc stars (Paper~III).

The existence of a dwarf so close to the Galactic disc, along with structures like the Mon Ring and the TriAnd
feature raises questions on their role in the formation of the Galactic (thick) disc (Paper~I). However, answering this
question first requires us to determine if there is a link between these three structures. \citet{penarrubia05} used all
known distance, radial velocity and proper motion data to constrain simulations of the
formation of the Mon Ring by an accretion event. Interestingly, they predict the progenitor of the Ring
should be located around the position of CMa ($l\sim245\deg$ and $b\sim-18\deg$) but at twice the estimated
distance of the CMa dwarf. Therefore, it remains unclear whether the two structures are linked or not.

This paper is the first in a series that aims at analyzing the kinematics of these different low latitude Galactic
structures. In particular, we aim at constraining the orbit of the accreted dwarf and determining if the Mon Ring could be
a by-product of this accretion. Here, we present our complete sample of radial velocities of Red Giant Branch and Red
Clump stars in the CMa overdensity and show it presents non-Galactic features. In section~2 we discuss the data set we
use for our radial velocity study of section~3. These radial velocities are used in section~4, to select CMa stars from
the UCAC2.0 catalogue and determine the proper motion of the dwarf. The obtained kinematics
are then used in section~5 to constrain simulations of the accretion of a dwarf onto the Galactic disc. Conclusions are
presented in section~6.

In the following, all the J, H, K magnitudes from 2MASS have been corrected from extinction using the maps from
\citet{schlegel98}, modified by the asymptotic correction from \citet{bonifacio00}. We also assume that the Solar radius
is $R_\odot = 8\kpc$, that the LSR circular velocity is $220\kms$, and that the peculiar motion of the Sun is
($U_0=10.00\kms, V_0=5.25\kms, W_0=7.17\kms$; \citealt{dehnen98}). Except when stated otherwise, the radial velocities
are the observed Heliocentric radial velocity, not corrected for the motion of the Sun.

\section{Observations and reduction}
The aim of the low Galactic latitude survey we undertook with the 2dF spectrograph was to determine the kinematics of
the different asymmetries that appear in the distribution of Red Giant Branch (RGB) stars (Figures~3 to 5 of
Paper~I) and in particular kinematics of the Canis Major dwarf galaxy. Hence, four fields out of the $\sim 15$
in our survey were directly aimed at the centre of the CMa overdensity as found in Paper~I:
$(l,b)=(240.0\deg,-4.8\deg)$, $(l,b)=(240.0\deg,-6.8\deg)$, $(l,b)=(240.0\deg,-8.8\deg)$ and
$(l,b)=(240.0\deg,-10.8\deg)$. We also observed another field on the opposite side of the Galactic disc, at
$(l,b)=(240.0\deg,+8.8\deg)$ to provide a reference to compare with the CMa fields.

\begin{figure}
\ifthenelse{\UseFigs=1}{
\includegraphics[angle=270, width=\hsize]{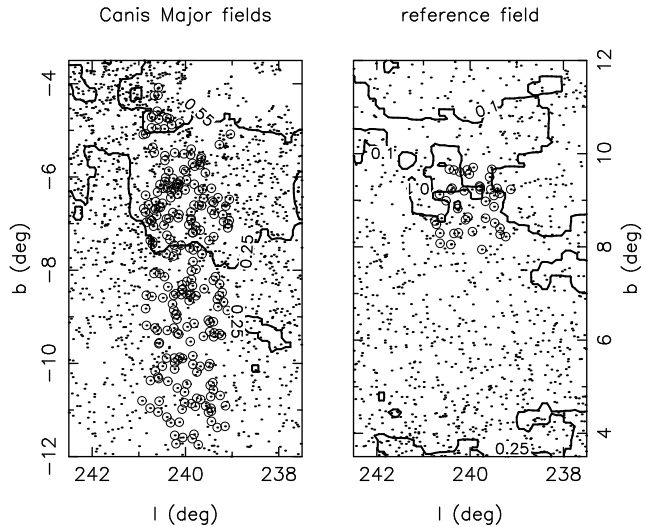} }{caption1}
\caption{Map of the target Red Giant Branch stars for the CMa region (left panel) and the reference field (right panel).
All RGB stars of these regions are shown as dots and target stars are circled. Extinction contours for $E(B-V)=0.1$,
0.25 and 0.55 are also shown as thick lines. Target Red Clump stars follow the distribution of target RGB stars but are
not shown to avoid overcrowding the plots.}
\end{figure}

In each two-degree field, our primary targets were the RGB stars selected from sample A of Paper~I and within 4 to
20~kiloparsecs from the Sun\footnote{We recall that the distance to RGB stars were determined by the photometric
parallax technique presented by \citet{majewski03} that we applied to a sample of CMa stars in Paper~III.}. These stars
are highlighted on the map of sample A stars shown on Figure~1 and mainly fall in regions of reasonable extinction
($E(B-V)<0.4$). Only the field closest to the Galactic disc with $-5.8\deg<b<-3.8\deg$ suffers from significant variable
extinction. The target stars are also shown on the CMD of each field in Figure~2. As the CMa structure has a prominent
Red Clump (RC, see Papers~I and II), we placed the remaining fibres to observe a sample of RC stars at the estimated
distance of CMa: $5\kpc<D_\odot<8\kpc$, assuming a distance of $7.2\kpc$ for the bulk of the overdensity (Paper~III) and
an absolute magnitude of  $M_{K}=-1.5\pm0.2$ for the RC population of CMa (Paper~II), independent of colour which was
chosen within the range $0.5<J-K<0.65$. The corresponding selection box has also been drawn on Figure~2. It can be seen
that even in fields with increasing extinction values (panels $c$ and $d$), the selection box remains centred on the Red
Clump zone, hinting at a good extinction correction for stars at this distance.

\begin{figure}
\ifthenelse{\UseFigs=1}{
\includegraphics[angle=270, width=\hsize]{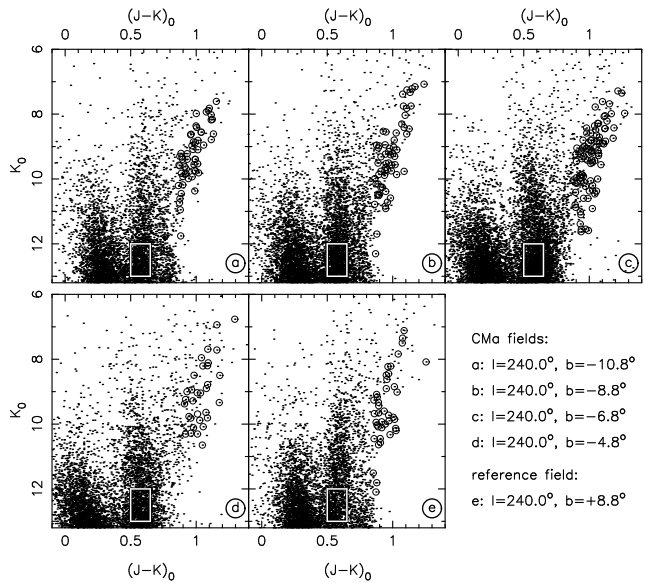} }{caption1}
\caption{2MASS infrared Colour Magnitude Diagrams of the four target fields around Canis Major (\textit{a} to
\textit{d} panels) and the symmetric field above the Galactic disc (panel \textit{e}). On each panel, the observed RGB
stars are shown as circled dots and the selection box from which Red Clump stars were chosen is drawn in white.}
\end{figure}

The observations were obtained during the nights of 7-12 April 2004. We employed two different spectrograph settings,
with the 1200V grating on spectrograph 1 (covering 4600--5600\AA\ at 1\AA/pixel) and with the 1200R grating on
spectrograph 2 (covering 8000--9000\AA, also at 1\AA/pixel). The observations have been reduced using the 2dF Data
Reduction package provided by the AAO \citep{taylor96} for the correction from the flat fields, the extraction of the
spectrum of each fibre and the sky subtraction.

\subsection{Reduction of first spectrograph spectra}
We use a custom-made reduction pipeline to correct the asymmetry of the Line Spread Function (LSF) of the 2dF with the
first spectrograph settings. This pipeline is described in detail in \citet{martin05} and can be summarized as follows: 

(i) the asymmetry of the LSF is modeled across the CCD for each observed field;

(ii) the spectrum of a given fibre is calibrated using the corresponding LSF model;

(iii) for each observed spectrum, the model is also used to generate template spectra. Hence, when doing a Fourier
cross-correlation to determine the radial velocity of an observed star, both the observed spectrum and the templates are
deformed in an identical way.

\noindent We have shown that this procedure ensures that we avoid systematic offsets higher than $\pm5\kms$ due to
deformations of the LSF.

\begin{table}
\begin{center}
\caption{Templates used for cross-correlation of first spectrograph observations.}
\begin{tabular}{lcc}
\hline\hline
Star      & spectral type & radial velocity ($\kms$)$^a$ \\
\hline
HD 145206 & K4III & -46.0\\
HD 167818 & K5III & -16.9\\
HD 149447 & K6III & -2.1\\
HD 89736 & K7III & 16.0\\
HD 92305 & M0III & -22.4\\
HD 102212 & M1III & 50.7\\
HD 120052 & M2III & 64.2\\
HD 224935 & M3III & -11.8\\
HD 11695 & M4III & 1.5\\
\hline
\end{tabular}
\end{center}
$^a$ Obtained using the SIMBAD database, operated at CDS, Strasbourg, France.\\
The high resolution spectra of these stars are extracted from the UVES Paranal Observatory Project \citep{bagnulo03}
\end{table}

Each observed spectrum was cross-correlated with up to 9 different templates of giant stars, from K4III to M4III
spectral types, generated from high resolution spectra from the UVES Paranal Observatory Project \citep[see
\citealt{martin05} and Table~1 for the list of templates]{bagnulo03}. RGB stars are cross-correlated with all nine
templates and have typical uncertainties of $\sim4-7\kms$ for each derived radial velocity while RC stars are
cross-correlated with the first five templates (from K4III to M0III spectral type), since their spectra are unlikely to
correspond to a higher spectral type, and have typical uncertainties of $\sim8-15\kms$ for each template. Instead of
choosing one of the template-specific derived radial velocities, we use a weighted average of the nine values to avoid
any systematic offset that could be due to particular features in one of the templates. Hence, the Heliocentric radial
velocity, $v_r$, of an observed star is given by:

\begin{equation}
v_r=\Big(\sum_{i=1}^{\mathrm{nb\,\,temp}}\frac{1}{\sigma_i^2}\Big)^{-1}\sum_{i=1}^{\mathrm{nb\,\,temp}}\frac{v_{r,i}}{\sigma_i^2}
\end{equation}

\noindent where $v_{r,i}$ is the radial velocity derived from the cross-correlation of the observed spectrum with the
$i$th template and $\sigma_i$ is the related uncertainty, given by the iraf function {\tt fxcor}. To judge the
homogeneity of the different $v_{r,i}$ around $v_r$, we also calculate the weighted dispersion, $\sigma_v'$, of the
$v_{r,i}$ around $v_r$:

\begin{equation}
\sigma_v'^2=\Big(\sum_{i=1}^{\mathrm{nb\,\,temp}}\frac{1}{\sigma_i^2}\Big)^{-1}\sum_{i=1}^{\mathrm{nb\,\,temp}}\frac{(v_r-v_{r,i})^2}{\sigma_i^2}.
\end{equation}

Stars with poorly determined radial velocities were eliminated by keeping only stars with $\sigma_v'<5\kms$ which
represent more than 90 percents of our sample.

\subsection{Reduction of second spectrograph spectra}

Only RC stars were observed with the second spectrograph settings. It has been shown in \citet{martin05} that these
settings produce no important deformation of the LSF and do not require the application of the reduction pipeline we
used for spectrograph 1 data. For each star, we fit a Gaussian model for each of the three lines of the Ca II triplet
and derive a velocity. We use the weighted average of these three velocity values as the radial velocity of the star and
compute the weighted dispersion $\sigma_v'$. Stars with $\sigma_v'<6.0\kms$ are kept as valid stars for the second
spectrograph settings (once again this represents more than 90 percent of the sample).

On both spectrographs, and to account for systematic offsets of no higher than $\sigma_{\mathrm{2dF}}=5\kms$ that may
remain after applying our reduction pipeline, we increase the uncertainty $\sigma_v'$ of each star to reach the total
uncertainty $\sigma_v$ on each radial velocities:

\begin{equation}
\sigma_v=\sqrt{\sigma_v'^2+\sigma_{\mathrm{2dF}}^2}
\end{equation}

Finally, we note that the observed spectra are of sufficiently good quality to avoid significant variation of
$\sigma_v$ with the magnitude of the target stars ($\sigma_v$ varies between 6 and $8\kms$ over the $7<K<13$ range of
our sample).

\section{Kinematic results}

\subsection{The properties of the RGB sample at the centre of CMa}

\begin{figure}
\ifthenelse{\UseFigs=1}{
\includegraphics[width=\hsize]{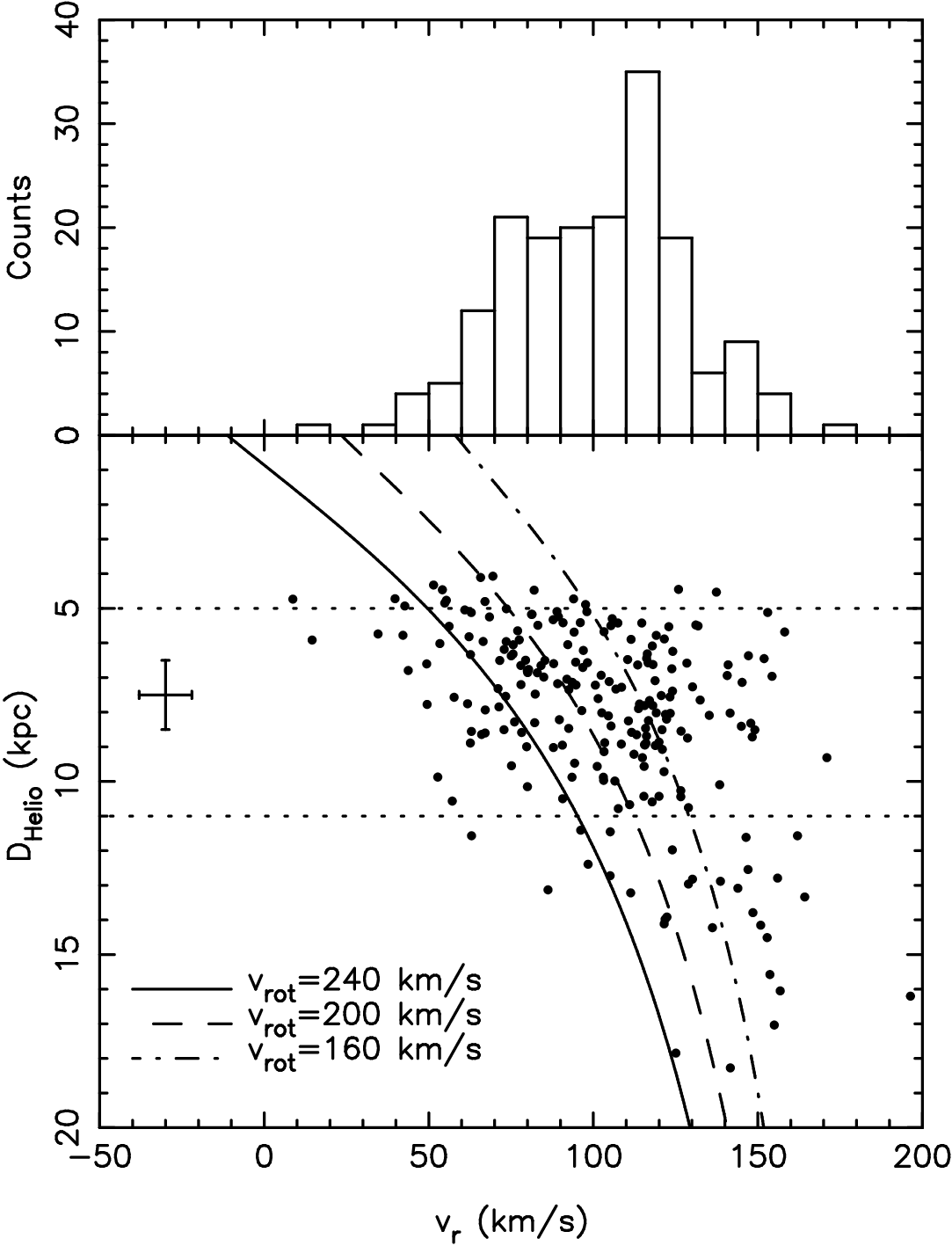} }{caption1}
\caption{The top panel shows the radial velocity distribution of RGB stars near the centre of CMa
($5\kpc<D_\odot<11\kpc$). A peak of stars is present at $\sim 115\kms$. The lower panel shows the position of
these RGB stars in phase space, with the typical error on both radial velocity and distance reported on the left. The
peak is produced by a group of stars clustered between $7\kpc$ and $9\kpc$. The expected position of a population
orbiting the Milky Way in a circular orbit at $v_{\mathrm{rot}}=240\kms$ (full line), $v_{\mathrm{rot}}=200\kms$ (dashed
line) and $v_{\mathrm{rot}}=160\kms$ (dashed-dotted line) has been overplotted for comparison. The stars selected to
produce the histogram of the upper panel are those between the two dotted line.}
\end{figure}

Our sample of RGB stars show no significant change with Galactic latitude hence, we merge the data of all the fields
with $b<0\deg$. The distribution of radial velocities of the 228 RGB stars in these four CMa fields is shown on the top
panel of Figure~3 selected around the estimated distance to the structure ($5\kpc<D_\odot<11\kpc$). The distribution is
rather broad, but has a well-defined peak of stars at $\sim115\kms$. 

\subsubsection{Comparison with the \citet{martin04b} results}
In Paper~III, we presented a similar analysis of the radial velocity of RGB stars near the centre of Canis Major based on
only the two central fields of the present survey [$(l,b)=(240.0\deg,-6.8\deg)$ and $(l,b)=(240.0\deg,-8.8\deg)$].
However, in that analysis the distribution of radial velocity showed two narrow peaks, centred on 60 and $110\kms$, that
are not reproduced here. The second peak is visible on Figure~3, but the first one has completely disappeared. This peak
was in fact artificially produced by template issues resulting from a fluctuating LSF asymmetry during the different
observation nights.

\begin{figure}
\ifthenelse{\UseFigs=1}{
\includegraphics[angle=270,width=\hsize]{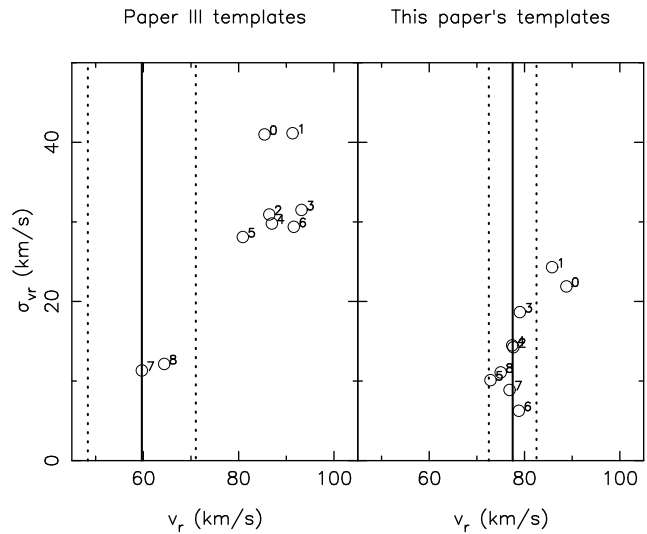} }{caption1}
\caption{Distribution of the radial velocities produced by the different template of Paper~III (left panel) and this
paper (right panel) for the same RGB star. The thick lines represent the radial velocity used in the two papers and
the dotted line represent the $\pm1\sigma_{v}$ limit. The reduction pipeline used here yields much more clustered values
than before. See the text for more details.}
\end{figure}

This is illustrated on Figure~4 where, for one of the stars belonging to the first peak of Paper~III, we present the
radial velocities derived from the different observed templates (left panel) and those derived using the artificial
templates to correct for the LSF asymmetry (right panel). Using our new reduction pipeline, most of the derived radial
velocities are well clustered around $v_r$ as defined in equation (1). Only two templates are over the 1-$\sigma$ limit
but they have high radial velocity uncertainties and hence, are not significant in the determination of the combined
$v_r$. On the other hand, the radial velocities derived for the different templates used in Paper~III are widely
scattered (over $35\kms$), and are not uniformly distributed. In particular, the two low uncertainty values are
clustered around $\sim60\kms$ while all the other values are clustered not far from the new radial velocity derived by
the pipeline. Since in Paper~III we only used the radial velocity that had the lowest uncertainty as the radial velocity
of the studied star, this star (as well as multiple other ones) ended up artificially populating a peak of stars at
$\sim60\kms$.

\subsubsection{The Canis Major radial velocity signature}
\begin{figure}
\ifthenelse{\UseFigs=1}{
\includegraphics[width=\hsize]{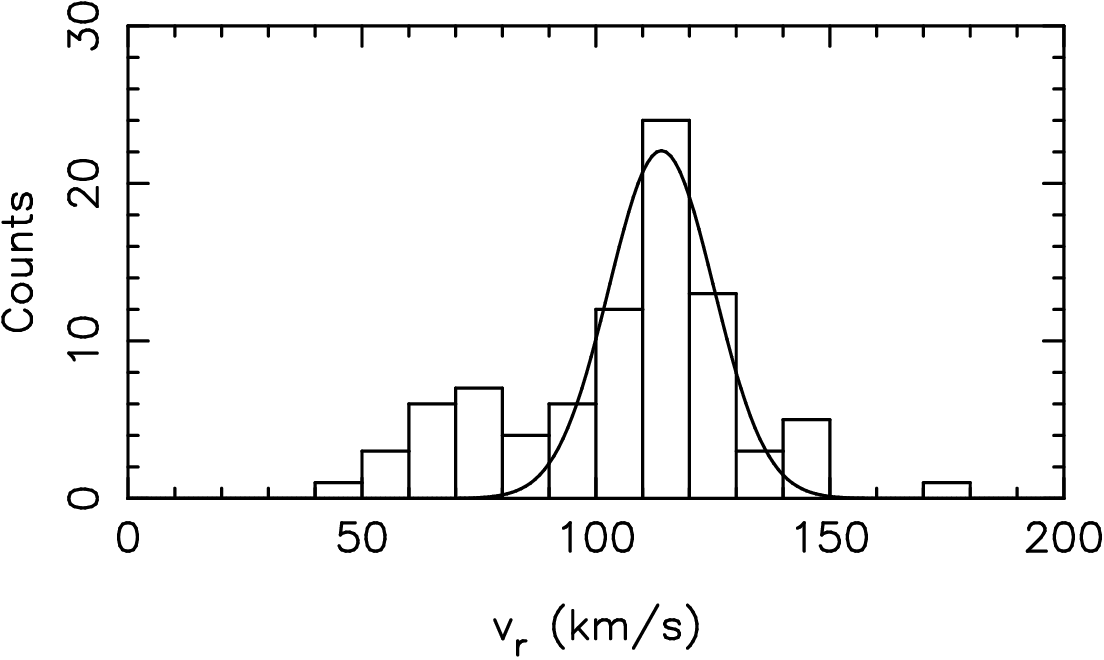} }{caption1}
\caption{The radial velocity distribution of those stars within 7.5 and $11\kpc$ from the Sun in the four CMa fields. 
The peak in the distribution is well reproduced by a Gaussian model centred on $114\pm2\kms$ and with an intrinsic
dispersion of $11_{-1}^{+3}\kms$.}
\end{figure}

Looking at the position of the RGB stars in phase space (bottom panel of Figure~3) reveals the peak in the
distribution is produced by a clump of stars with $7.5\kpc<D_\odot<11.0\kpc$. A more detailed analysis of this distance
range (Figure~5) reveals a population with a narrow dispersion. Comparing the data between $85\kms$ and $145\kms$ with a
Gaussian model using a maximum likelihood technique reveals this population is centred on $114\pm2\kms$ and has an
intrinsic dispersion of $11_{-1}^{+3}\kms$ when accounting for radial velocity uncertainties as in equation (3). As we
explained previously in Paper~III, the radial velocity of this population is not easily compatible with a disc-like
population since it would be orbiting the Galaxy with a rotational velocity of only $\sim160\kms$. Given the distance
and peculiar velocity of this population, we assume that it corresponds to the CMa dwarf. Changing the lower distance
cut in the range $7\kpc$ to $8\kpc$ does not substantially modify these results (position and dispersion change by less
than $2\kms$).

At closer distance, even though 2MASS starcounts of RGB stars betray the presence of the CMa dwarf (Paper~III), most
of the stars follow a more disc-like velocity. Yet, a group of these stars seems aligned in phase space, from
$D_\odot\sim6\kpc$ and $v_r\sim70\kms$ to the position of the clump of stars at higher distance. The low number of RGB
stars in our sample prevents any firm conclusion on the reality of this feature, even though it seems to have a low
dispersion. In the same distance range, a more diffuse group of stars also appears at higher radial velocity
($v_r>140\kms$). Given the relatively sparse nature of these stars they could simply be the tail of the distribution
of disc stars and/or could possibly represent a contamination from halo stars along these sight lines.

Another interesting feature of the phase space diagram is the group of stars at high distances (mainly
$11\kpc<D_\odot<15\kpc$ corresponding to $16.5\kpc<D_{GC}<20\kpc$) that seems to be disconnected from the CMa population
by a gap at $D_\odot\sim11\kpc$. The Galactic disc is known to possess a cut off at $D_{GC}\sim15\kpc$
\citep[e.g.][]{ruphy96} and hence, no disc stars are expected at the distance of this group. Moreover, given the high
radial velocity of these stars, it seems unlikely that this group is composed of misidentified dwarfs. In fact, since
these stars appear to be located between 16.5 to $20\kpc$ from the Galactic centre we believe we have uncovered part of
the Monoceros Ring behind the CMa dwarf. The radial velocity signature of this population is compatible with previous
work but we defer a more thorough analysis of the presence of the Monoceros Ring in all our low latitude fields to
another contribution (Conn et al. in preparation).

\subsection{The RC sample at the centre of CMa}

\begin{figure}
\ifthenelse{\UseFigs=1}{
\includegraphics[width=\hsize]{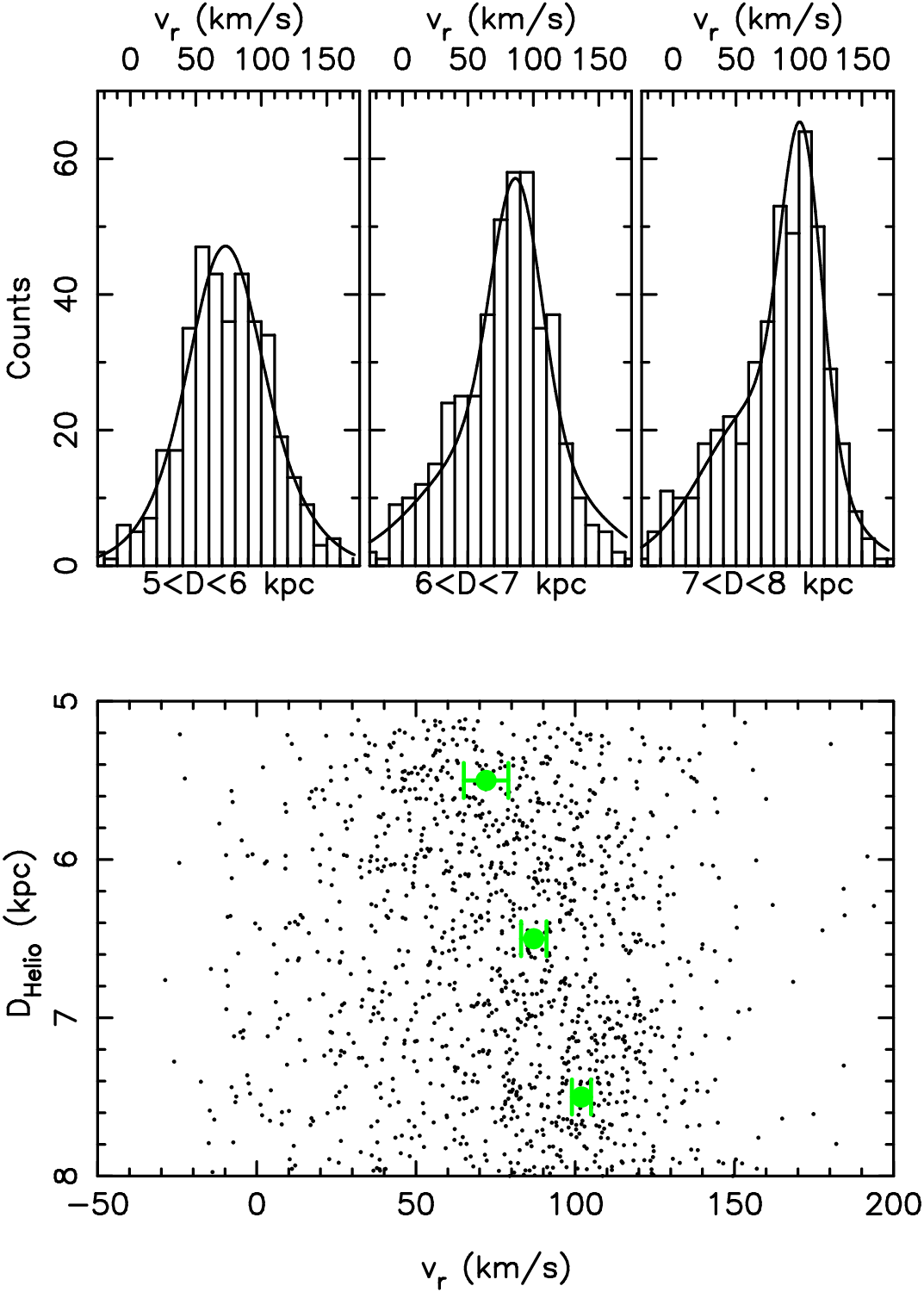} }{caption1}
\caption{Phase space distribution of stars in the RC sample (bottom panel). A significant part of the stars follows a
distance -- radial velocity relation that is confirmed when analyzing the radial velocity distribution of stars in
$1\kpc$ bins (histograms of top panels, with the fits corresponding to the values of Table~2). The positions of this
peculiar population have been overplotted as filled circles in the bottom panel, with error bars representing
$1\sigma$ uncertainties on these values.}
\end{figure}

Most of the RC stars were observed in the $(l,b)=(240.0\deg,-8.8\deg)$ field and should give a more precise view of 
the kinematics at the core of the CMa dwarf. While restricted to a shorter range in distance, the RC sample has the
advantage of containing $\sim1350$ stars at the distance of the group of RGB stars that seem to be aligned in phase
space. The distribution of these RC stars in phase space (bottom panel of Figure~6) presents a behaviour that is very
similar to the RGB sample, with an overdensity of stars that follow a distance -- radial velocity relation. The
distributions of the radial velocity of these stars for 1\kpc\ bins are shown on the top panels of Figure~6. We
suspect two populations are present in each bin: contaminating stars that fall in our CMD selection box of RC stars
and genuine Canis Major RC stars. To determine whether it is worth fitting a double Gaussian model to the distribution, we
use the {\sc kmm} test presented in \citet{ashman94}. The probabilities $P$ that the data are better represented by a
single Gaussian instead of two are $P(5<D_\odot<6)=1.0$, $P(6<D_\odot<7)=6\cdot10^{-3}$ and $P(7<D_\odot<8)<10^{-3}$ for
the three distance bins. The last two bins are hence inconsistent with a single Gaussian model and
only the closest one would be better characterized by a single Gaussian. However, since the distance -- radial velocity
relation still seems to be present in this bin, we believe that the two populations appear in this bin but that their mean
velocity is located at around the same position. Hence, we proceed in fitting a double Gaussian model to the data in
all bins, knowing that uncertainties on the derived values for the first bin will be large.

\begin{table}
\begin{center}
\caption{Parameters of the Red Clump radial velocity distribution fits.}
\begin{tabular}{lcccc}
\hline\hline
Distance bin      & $<v_{r,\mathrm{CMa}}>$ & $\sigma_{\mathrm{CMa}}$  & $<v_{r,\mathrm{cont}}>$ &
$\sigma_{\mathrm{cont}}$\\
(kpc)                & $(\kms)$ & $(\kms)$ & $(\kms)$  & $(\kms)$ \\
\hline
$5<D_\odot<6$   & $72\pm7$ & $24\pm4$ & $75\pm6$ & $42\pm4$\\
$6<D_\odot<7$   & $87\pm4$ & $20\pm3$ & $76\pm5$ & $54\pm4$\\
$7<D_\odot<8$   & $102\pm3$ & $16\pm3$ & $71\pm5$ & $46\pm3$\\
\hline
\end{tabular}
\end{center}
Column 1 and 2 state, respectively, the mean velocity and the intrinsic dispersion of the CMa population. Columns 3 and 4
list the same parameters for the contaminating population.
\end{table}

We use a maximum likelihood technique to fit the double Gaussian model. The uncertainties $\sigma_{v}$ on each radial
velocity were taken into account whereas no uncertainty was assumed for RC distance since the only important source of
error comes from the absolute magnitude of the population and a change in this value would only shift the sample as a
whole. Since the two populations are most clearly separated in the $7<D_{\odot}<8\kpc$, we use the stars in this bin
to determine the proportion of RC stars that belong to the Canis Major population. The fit yields to a 40 percent
proportion of CMa stars, well within the 25 to 50 percent determined by \citet{bellazzini04}. Since we expect this
proportion to be roughly constant over our 3\kpc\ sample (see Figure~2 of Paper~II), we also adopt this value for the two
closest bins. The four parameters of the fits -- mean velocity $<v_{r,\mathrm{CMa}}>$ and intrinsic dispersion
$\sigma_{\mathrm{CMa}}$ of the Canis Major population and mean velocity $<v_{r,\mathrm{cont}}>$ and intrinsic dispersion
$\sigma_{\mathrm{cont}}$ of the contaminating population -- are presented in Table~2.

Each distance bin contains a low dispersion population with an increasing mean radial velocity and a broad population of
constant mean velocity. This latter population shows a velocity dispersion $\sigma_{\mathrm{cont}}$ comparable to Solar
neighbourhood velocity dispersions of the thick disc $(\sigma_U,\sigma_V, \sigma_W)=(63\pm6, 39\pm4, 39\pm4)\kms$
\citep{soubiran03}. We believe these stars are disc stars and dwarfs close to the Sun that fall in the RC region of the
(J-K,K) CMD and contaminate our sample. On the other hand, the other population that we believe to be composed of Canis
Major stars is kinematically cold, with an intrinsic internal dispersion around $16\pm3\kms$. It is likely that the
increase in $\sigma_{\mathrm{CMa}}$ in the other distance bins is only due to difficulties for the fit to disentangle
the two populations. The other notable feature of these stars is their increasing radial velocity with distance, from
$72\pm7\kms$ to $102\pm3\kms$ over only 3\kpc. Comparison with the group of RGB stars we identified as Canis Major stars
at higher distance reveals striking similarities with both RC and RGB samples being kinematically cold and having
compatible intrinsic dispersion. Moreover, the distance -- radial velocity trend that appears in the RC sample is nicely
prolonged to higher distances by the group of Canis Major RGB stars at $D_{\odot}\sim8.5\kpc$ and overlap with the
group of RGB stars that seem to be align in phase space for the $D_{\odot}>6\kpc$ region. It is
unlikely that a disc-like population could show characteristics that are similar to these stars given the high shift in
radial velocity over only a few kiloparsecs. Indeed, if the considered population was rotating
around the Milky Way on a circular orbit, it would have a rotational velocity as high as $v_{rot}\sim220\kms$ at
$5.5\kpc$ and as low as $v_{rot}\sim160\kms$ at $8.5\kpc$. On the other hand, correlation in phase space and low
dispersion in radial velocity, two characteristics that the CMa population shows, are typical of an accretion process
\citep[see e.g.][for the Sgr dwarf]{ibata97}.

\begin{figure}
\ifthenelse{\UseFigs=1}{
\includegraphics[width=\hsize]{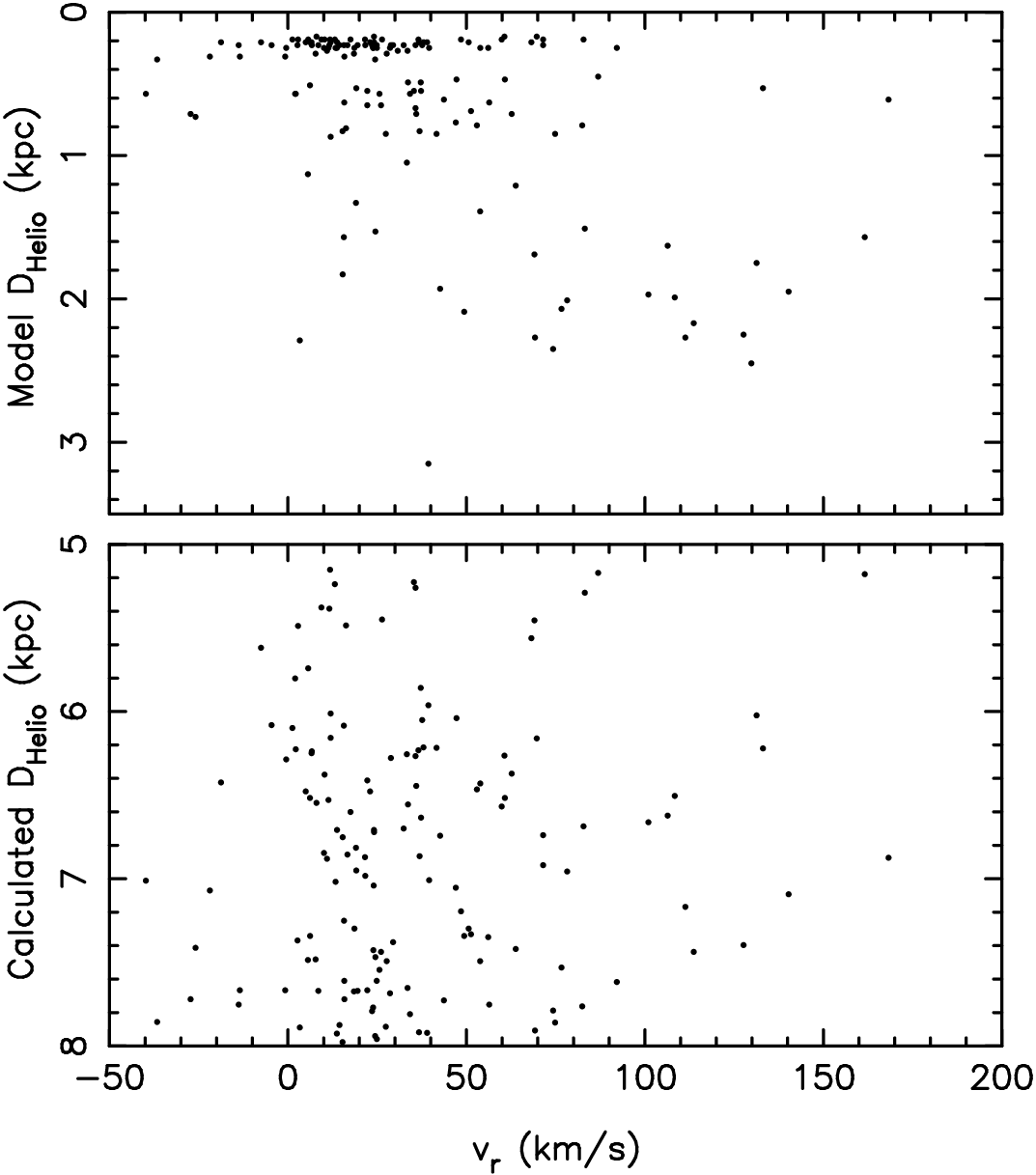} }{caption1}
\caption{Phase space distribution of stars in the Besan\c{c}on model that fall in our selection box for RC stars in the
region of the observed 2dF fields. For the top panel, the distance of the stars in the model was used and shows that 
most of the contamination in the observed sample is due to stars close to the Sun. For the bottom panel, we calculate the
distance as we do for the observed sample of RC stars. A direct comparison of this panel with Figure~6 shows that the
distance -- radial velocity relation in our sample is not predicted by the model.}
\end{figure}

To further test this conclusion, we compare the observations with the Besan\c{c}on model \citep{robin03}. We simulate
the J,H,K Colour-Magnitude Diagram of the region $239\deg<l<241\deg$ and $-4\deg<b<-12\deg$, from which we extract Red
Clump stars with the same colour cuts as those we used to define our RC sample. We then calculate the distance to the
stars in the model as we do for the observed stars, assuming an absolute magnitude of $M_K=-1.5$. Of
course, since these stars do not belong to the CMa dwarf, their estimated distance is different from their distance
provided by the model. From the $\sim150$ stars in the model for the selected region, four fifths are in fact
located at less than $1\kpc$ from the Sun. The distribution of these stars in phase space (Figure~7) therefore represents
that of the close disc stars that contaminate our sample of CMa RC stars. They do not follow the distance -- radial
velocity relation that we attributed to the CMa dwarf and are, on the contrary, located at a lower velocity. A similar
behaviour is observed for RGB stars in the model.

\subsection{The $(l,b)=(240.0\deg,+8.8\deg)$ field}

\begin{figure}
\ifthenelse{\UseFigs=1}{
\includegraphics[width=\hsize]{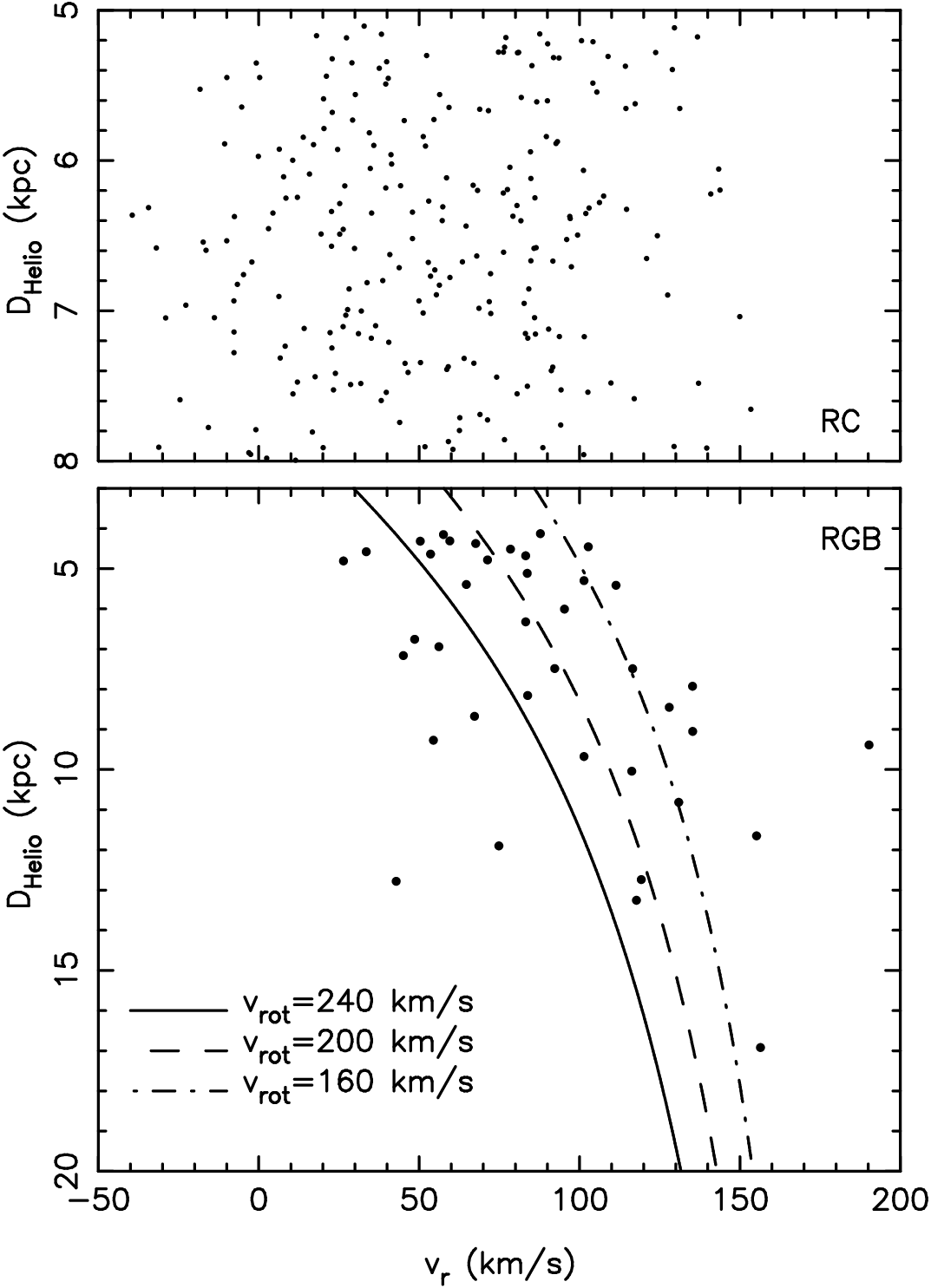} }{caption1}
\caption{Phase space distribution of RC (top panel) and RGB stars (bottom panel) in the
$(l,b)=(240.0\deg,+8.8\deg)$ field. Neither of the two show the features that are present in fields centred on the CMa
dwarf galaxy.}
\end{figure}

Another check of the peculiar radial velocity of the stars in the CMa structure is given by the field that is symmetric
to the CMa fields on the other side of the Galactic disc and that does not show the features that appear at the centre
of the CMa dwarf (Figure~8). The RGB sample for this field contains stars that are mostly located at $D_\odot<6\kpc$
with the expected radial velocity of a population orbiting the Milky Way at $v_{\mathrm{rot}}\sim200\kms$.
Only 11 stars fall within the distance cut that was used to produce Figure~5 near the centre of CMa.
Moreover, the remaining stars at higher distances show no sign of clustering. The RC stars in this field follow a
broad distribution with a dispersion of $\sim50\kms$, centred around $\sim60 \kms$ over the 3\kpc\ sample, similar to
what we identify as the contaminating population in the CMa field, and also similar to what would be expected from close
stars artificially spread over the 3\kpc\ distance range. Therefore, the peculiar features that appear at the centre of
CMa have to be due to the CMa population.

\section{UCAC2.0  proper motions of the CMa dwarf}

\begin{figure}
\ifthenelse{\UseFigs=1}{
\includegraphics[angle=270,width=\hsize]{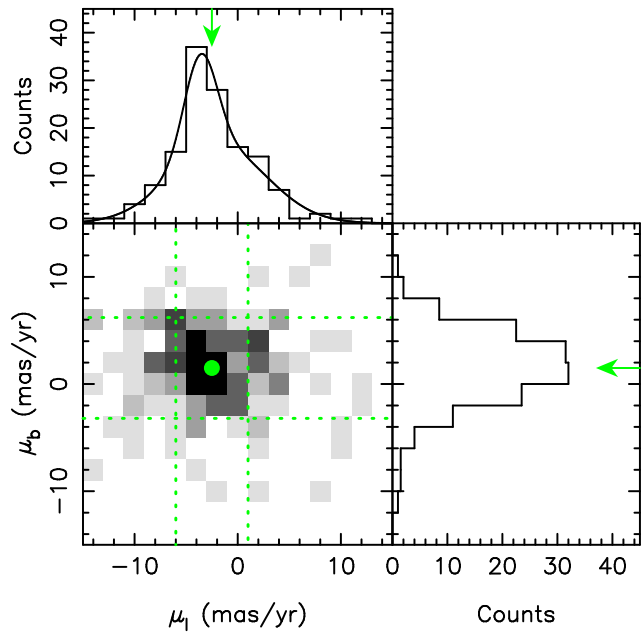} }{caption1}
\caption{The bottom left panel shows the distribution in proper motion space of CMa RC stars selected along the distance
-- radial velocity feature of Figure~6. A black pixel corresponds to 8 stars. The centre of the distribution (assuming
Gaussian distributions) is overplotted as a dot. The distribution of those stars within 1 $\sigma$ of this point in
the $\mu_b$ (respectively $\mu_l$) direction where used to produce the distribution of CMa stars in $\mu_l$ (resp.
$\mu_b$) on the top (resp. right) panel. The mean position derived from the Gaussian fits of the whole distribution are
shown as arrows. A wing in the distribution in $\mu_l$ betrays the presence of two populations that we fit by a double
Gaussian model.}
\end{figure}

\citet{momany04} used the UCAC2.0 catalogue \citep{zacharias04} to show that the CMa overdensity of RGB stars rotates
around the Galaxy in a prograde motion. However, given the sizable uncertainties on the proper motion values of
individual stars and the contamination of disc stars in the sample, they were only able to give a rough estimate of the
proper motions that convert in Galactic coordinates as: $\mu_l=-3.5\masyr$ and $\mu_b=-0.1\masyr$ with uncertainties of
$\sim2.0\masyr$.

With the addition of radial velocities of stars in the direction of CMa, we are now in a position to define a less
contaminated sample of stars that belong to the dwarf. We first determine the best linear fit to the distance -- radial
velocity relation of the RC sample by using the three radial velocity measurements for the three distance bins. We then
extract from  UCAC2.0 the proper motions of the RC stars that are located within $\pm10\kms$ of this linear fit. This
selection should ensure that a sizable proportion of our objects indeed belong to the CMa dwarf. To account for the
distance $D_\odot$ of the different stars, we normalize each proper motion $\mu_{\mathrm{UCAC}}$ as if the star was
located at the Heliocentric distance of the CMa dwarf ($D_{\odot,\mathrm{CMa}}=7.2\kpc$):

\begin{equation}
\mu=\mu_{\mathrm{UCAC}}\cdot\frac{D_\odot}{D_{\odot,\mathrm{CMa}}}
\end{equation}

\noindent The distribution of these $\mu$ proper motions is shown on the bottom left panel of Figure~9 and is consistent
with the \citet{momany04} measurements.

We use a maximum likelihood technique to determine the best fit of this distribution by a two dimensional Gaussian
function. The best mean value is $(\overline{\mu_l},\overline{\mu_b})=(-2.5\masyr,1.5\masyr)$ with internal dispersions
of $(\sigma_{\mu l},\sigma_{\mu b})=(4.8\masyr,3.7\masyr)$. To analyze in more depth the proper motions of the CMa
population, we use these values to eliminate outliers and study the $\mu_l$ distribution of stars within $\sigma_{\mu b}$
of $\overline{\mu_b}$ (top panel of Figure~9) and the $\mu_b$ distribution of stars within $\sigma_{\mu l}$ of
$\overline{\mu_l}$ (right panel of Figure~9). This latter distribution is well centred around $\overline{\mu_b}$ (the
arrow) and the value $1.5\pm0.5\masyr$ can be taken as the proper motion of the CMa dwarf in Galactic latitude. On the
other hand, the distribution of proper motions in Galactic longitude is not centred around $\overline{\mu_l}$. A peak appears
at lower proper motion, with a wing that extends to positive values. Considering that, even with our cut in radial
velocity, contaminating stars from the disc should still be in our sample of stars, we use a maximum likelihood
technique to fit a double Gaussian model to the distribution (the {\sc kmm} test yields a low probability of $10^{-2}$
that the population is better represented by a single Gaussian). This reveals two distinct populations are present:
similar to what was observed in radial velocity, one has a large dispersion ($4.7\masyr$), is centred on
$\mu_l=-1.9\masyr$ and accounts for one fourth of the total number of stars while the other has a much narrower
dispersion ($1.5\pm0.4\masyr$) and is centred on $\mu_l=-3.6\pm0.8\masyr$. Since one expects the contaminating stars in
the sample to be at different distances and hence have different proper motions, we believe the broad population
corresponds to these stars; so the narrow population has to be due to the CMa dwarf.

\begin{figure}
\ifthenelse{\UseFigs=1}{
\includegraphics[angle=270,width=\hsize]{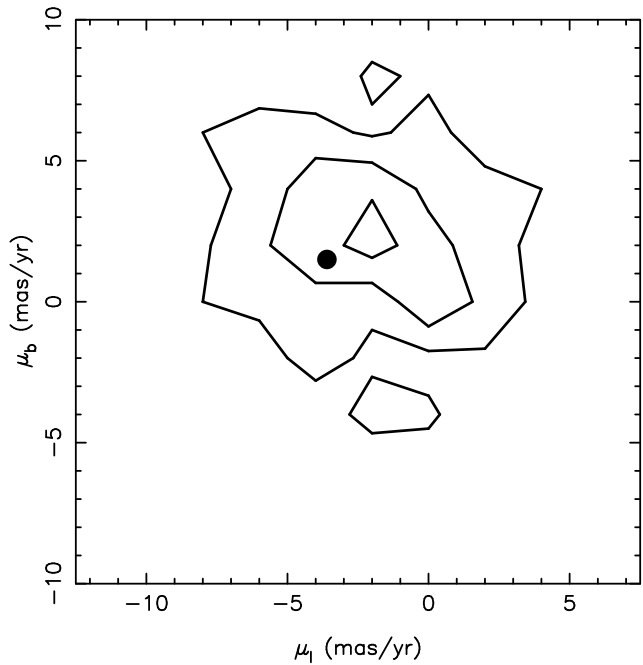} }{caption1}
\caption{The contours show the distribution of RC stars in our sample that are more the $20\kms$ away from the
distance -- radial velocity relation followed by CMa stars. This population is centred around
$(\mu_{l,cont},\mu_{b,cont})\sim(-2.0\masyr,2.0\masyr)$ and offset from the CMa population whose mean proper motion
is represented by the black dot.}
\end{figure}

For comparison, we show on Figure~10 the distribution of RC stars in our sample that are more than $\pm20\kms$ away
from the linear fit of the Canis Major radial velocity and that should correspond to the contaminating population in the
sample. This population has a proper motion of $(\mu_{l,\mathrm{cont}},\mu_{b,\mathrm{cont}})\sim(-2.0\masyr,2.0\masyr)$
offset from the CMa proper motion, especially for the proper motion in Galactic longitude.

The derived proper motion values for the CMa dwarf $(\mu_l,\mu_b)=(-3.6\pm0.8\masyr,1.5\pm0.4\masyr)$ are
compatible with the \citet{momany04} values but with much lower uncertainties. Correcting from the reflex solar motion
along the line of sight yields $(\mu_l',\mu_b')= (-6.8\pm0.8\masyr, 0.8\pm0.4\masyr)$, corresponding to a tangential
velocity of $234\kms$ at the mean distance of 7.2\kpc.

\section{Simulations}

The kinematic information on the CMa dwarf provides interesting constraints on the accretion process that the dwarf is
undergoing. In this section, we revisit the simulations of Paper~I, this time using only constraints that
can be directly linked to CMa. Indeed, since there is no definite proof of a link between CMa and the Mon Ring, we choose
not to use the numerous kinematic and/or positional data on the Mon Ring for our simulations\footnote{Moreover, the
existence of the Mon Ring behind CMa seems to hint against a direct link between the two structures and it would be the
tidal arms of the dwarf, wrapped a few times around the Galaxy, that would produce the ring-like structure. If such a
scenario is plausible \citep[see e.g.][]{penarrubia05}, the constraints provided by the Ring data on the accretion
parameters of the dwarf are weak.}.

\begin{figure}
\ifthenelse{\UseFigs=1}{
\includegraphics[angle=270,width=\hsize]{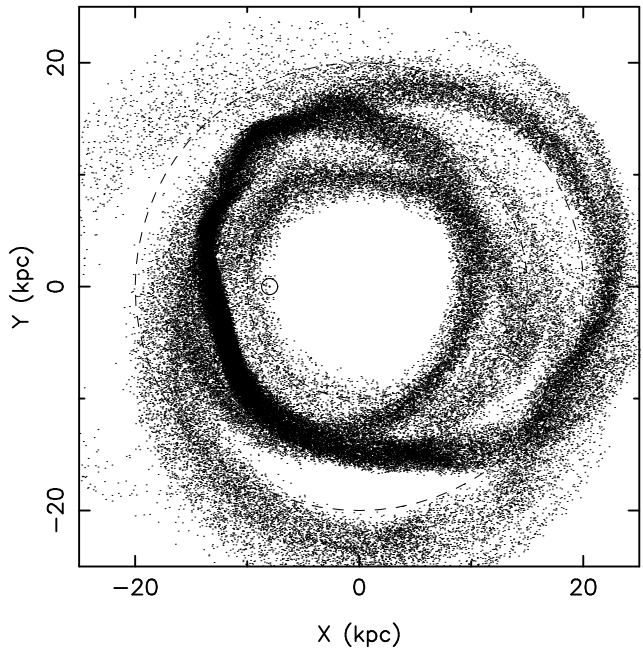} }{caption1}
\caption{Our best simulation of the accretion of the CMa dwarf viewed from the North Galactic Pole. The Sun is
represented by a dotted circle and the three dashed circles represent distances of 10, 15 and $20\kpc$ from the Galactic
centre. The main body of the CMa dwarf is visible at $(X,Y)\sim(-12\kpc,-6\kpc)$ and the tidal arms produced by the
accretion are wrapped multiple times around the Milky Way.}
\end{figure}

The simulations are performed in the same way as in \citet{martin04a}, using the fast and momentum-conserving tree code
integrator {\sc falcON} \citep{dehnen00,dehnen02}.  The simulation that best fits the kinematic data at the centre of
CMa is shown from the North Galactic Pole on Figure~11. It is produced by a dwarf galaxy modeled by a King model of
$5\times10^8\msun$ with a tidal radius of $2.7\kpc$ and $W_0=3.25$ and that is accreted onto the Milky Way during
$\sim3$~Gyr. The simulation reproduces at the same time the overdensity of stars that revealed the dwarf in 2MASS, the
proper motions we measured in this paper (with values of $\mu_l\sim-4\masyr$ and $\mu_b\sim2\masyr$) and, above all, the
distance -- radial velocity gradient that appears in RC (and possibly RGB) stars. This is shown on Figure~12, with the
values measured in this paper plotted as squares and triangle on top of the particles of the simulation in the region of
our 2dF fields. Moreover, the velocity dispersion of the particles reproduces that measured for the RGB sample
($11\kms$). Of course, the parameters of the initial dwarf are certainly not the only ones that can reproduce
the radial velocities at the centre of the dwarf but it should be noticed that the simulated position and velocities are
very near what is observed.

\begin{figure}
\ifthenelse{\UseFigs=1}{
\includegraphics[angle=270,width=\hsize]{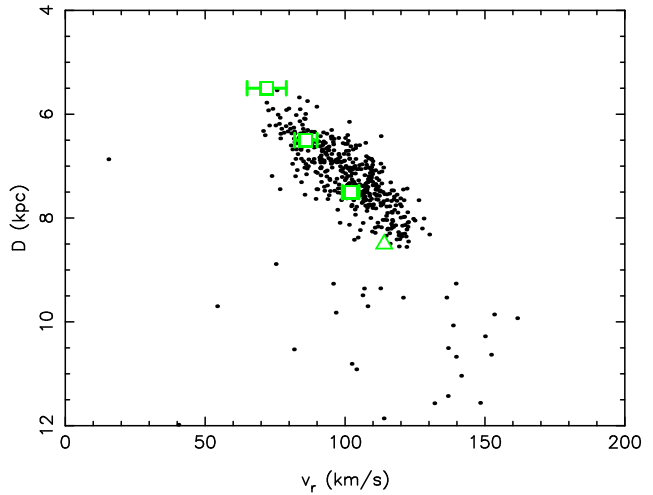} }{caption1}
\caption{Distribution in space phase of the particles (dots) of our best simulation at the location of the 2dF fields we
present in this work ($239\deg<l<241\deg$ and $-12\deg<b<-4\deg$). The observed values of radial velocity are shown as
squares for values derived from the RC sample and a triangle for the one derived from the RGB sample. They are 
well matched by the particles. The group of star at higher distances is produced by the multiple wraps of the tidal arms
of the dwarf around the Milky Way.}
\end{figure}

The group of particles that is visible at higher distance ($9\kpc<D<12\kpc$) is produced by one of the tidal arms of
the dwarf that is wrapped a few times around the Milky Way. Even if the group is not exactly at the same distance as
what we identify as the Mon ring behind the CMa dwarf, it is tempting to explain the two observations by the same
accretion process. It is also interesting to notice that the average location of these tidal arms in Figure~11 are
close to the distance range of the Mon ring ($10\kpc<D_{GC}<18\kpc$ in the anticentre direction) while the radial
velocity of the arm reveals it is only shifted by $\sim30\kms$ compared to the SDSS radial velocities of
\citet{yanny03}. The more distant portions of the arms, that extend at 20-30\kpc\ from the Galactic centre could also
explain the Triangulum-Andromeda structure presented in \citet{rocha-pinto04}. Yet, given the lack of definitive proof of
a link between the two structures, we chose not to try to fit the radial velocity of the group of Mon ring stars behind
CMa. Indeed, a small shift of $\sim10\kms$ in the tangential velocities of our simulation substantially changes the
position and kinematics of the tidal arms. Only with a determination of the radial velocity of the arms away from the
main body of the dwarf could such a constraint be useful.

\section{Conclusion}

We have shown that the Red Giant Branch and Red Clump stars at the centre of the CMa dwarf have peculiar kinematics
that are incompatible with a disc-like population:

(i) The phase space distribution of stars (whether RGB or RC) in the CMa fields under the Galactic disc (Figures~3
and~6) is very different from that of the symmetric region on the other side of the disc (Figure~8).

(ii) More than one third of the stars in the RC sample follow a distance -- radial velocity relation over all the
sample ($5\kpc<D_{\odot}<8\kpc$). This relation seems also to be present in the RGB sample and prolonged to higher
distances by a group of RGB stars. The important radial velocity shift over only a few kiloparsecs for these stars (which
is not reproduced by the Besan\c{c}on Galactic model) cannot be explained by a group of stars rotating around the Milky
Way on a circular orbit.

(iii) The proper motions of the dwarf that we derived from the UCAC2.0 catalogue
$(\mu_l,\mu_b)=(-3.6\pm0.8\masyr, 1.5\pm0.4\masyr)$ yields at tangential velocities when corrected from the
Solar motion of $(v_l,v_b)=(-235\pm35\kms,+15\pm25\kms)$. Combined with a mean radial velocity of
$v_r\sim-105\kms$ (also corrected from the Solar motion), the total velocity of the CMa population is
$v_{tot}\sim260\kms$ which is high compared to expectations for disc stars \citep[e.g.][]{soubiran03}.

(iv) Both the RGB and RC samples show the CMa population is kinematically cold, with intrinsic dispersions of 11 and
$16\kms$ respectively but this latter value may be overestimated due to the difficulty of separating CMa stars from
contaminating stars in the RC sample. Such low dispersions are typical of recent accretion process and can be
observed, for instance, in the Sagittarius dwarf galaxy \citep{ibata97}.

(v) Following the discovery of a distance spiral arm at the edge of the Galactic disc by \citet{mcclure04}, we mentioned
in Paper~III, that the radial velocity of the Canis Major RGB population could be compatible with this spiral arm.
However, the distance -- radial velocity of RC CMa stars reach a much lower value of $72\kms$ hardly compatible with
the \citet{mcclure04} values. Moreover, {\sc Hi} maps do not show any hint of a spiral arm at the distance of the Canis
Major object \citep[see e.g.][]{nakanishi03}.

All these observations point at the accretion scenario for the CMa population, with a dwarf galaxy that is currently
being absorbed by the Milky Way. The low velocity dispersion suggests that, even though it is undergoing dramatic tidal
stripping by our Galaxy, the dwarf is still bound. But can the CMa dwarf be also responsible for the Monoceros ring
and the Triangulum-Andromeda structure? If simulations can reproduce the observed kinematic signature of CMa,
constraints on the tidal arms of the dwarf are at the moment not strong enough to definitely conclude on the CMa dwarf
being the progenitor of the Mon ring.

A definite conclusion may only come with future observations which should concentrate on mapping the regions around the
centre of dwarf ($220\deg\simlt l\simlt 260\deg$ and $-20\deg<b<0\deg$). Moreover, in addition to constraining the orbit
of the accretion, determining the intrinsic dispersion of the CMa population and the evolution of this dispersion around
its core should allow a much more precise estimate of the parameters of the initial dwarf, as well as an estimate of its
mass-loss rate \citep{johnston99}.

\section*{Acknowledgments}
The referee is thanked for many useful comments that helped improve the overall quality of the paper.

\newcommand{\mnras}{MNRAS}
\newcommand{\pasa}{PASA}
\newcommand{\nat}{Nature}
\newcommand{\araa}{ARAA}
\newcommand{\aj}{AJ}
\newcommand{\apj}{ApJ}
\newcommand{\apjl}{ApJ}
\newcommand{\apjs}{ApJSupp}
\newcommand{\aap}{A\&A}
\newcommand{\aaps}{A\&ASupp}
\newcommand{\pasp}{PASP}


\begin{thebibliography}{}
%
\bibitem[Ashman, Bird \& Zepf(1994)]{ashman94}
        Ashman K. M., Bird C. M. \& Zepf S. E. 1994, \aj\ 108, 2348
%
\bibitem[Bagnulo et al.(2003)]{bagnulo03}
        Bagnulo, S., Jehin, E., Ledoux, C., Cabanac, R., Melo, C., Gilmozzi, R., The ESO Paranal Science Operations Team
2003, The Messenger, 114, 10
%
\bibitem[Bellazzini et al.(2004)]{bellazzini04}
        Bellazzini M., Ibata R. A., Monaco L., Martin N. F., Irwin M. J. \& Lewis G. F. 2004, \mnras\ 354, 1263
%
\bibitem[Bonifacio, Monai \& Beers(2000)]{bonifacio00} 
        Bonifacio P., Monai S. \& Beers T. 2000, \aj\ 120, 2065
%
\bibitem[Crane et al.(2003)]{crane03}
        Crane J., Majewski S., Rocha-Pinto H., Frinchaboy P., Skrutskie M. \& Law D. 2003, \apj\ 594, L119
%
\bibitem[Conn et al.(2005)]{conn05}
        Conn B. C., Lewis G. F., Irwin M. J., Ibata R. A., Irwin J. M., Ferguson A. \& Tanvir N., \mnras\ submitted
%
\bibitem[Dehnen \& Binney(1998)]{dehnen98}
        Dehnen W. \& Binney J. 1998, \mnras\ 298, 387
%
\bibitem[Dehnen(2000)]{dehnen00}
        Dehnen W. (2000), \apj\ 536, 39L
%
\bibitem[Dehnen(2002)]{dehnen02}
        Dehnen W. (2002), J. Comput. Phys. 179, 27
%
\bibitem[Ibata et al.(1997)]{ibata97}
        Ibata R., Wyse R., Gilmore G., Irwin M. \& Suntzeff N. 1997, \aj\ 113, 634
%
\bibitem[Ibata et al.(2003)]{ibata03}
        Ibata R., Irwin M., Lewis G., Ferguson A. \& Tanvir N. 2003, \mnras\ 340, 21
%
\bibitem[Johnston, Sigurdsson \& Hernquist(1999)]{johnston99}
       Johnston K. V., Sigurdsson S. \& Hernquist L. 1999, \mnras\ 302, 771
%
\bibitem[Majewski et al.(2003)]{majewski03}
        Majewski S., Skrutskie M., Weinberg M. \& Ostheimer J. 2003, \apj\ 599, 1082
%
\bibitem[Martin et al.(2004a)]{martin04a}
        Martin N. F., Ibata R. A., Bellazzini M., Irwin M. J., Lewis G. F. \& Dehnen W. 2004a, \mnras\ 348, 12 (Paper~I)
%
\bibitem[Martin et al.(2004b)]{martin04b}
        Martin N. F., Ibata R. A., Conn B. C., Lewis G. F., Bellazzini M., Irwin M. J.,\& McConnachie A. W. 2004b,
\mnras\ 355, L33
%
\bibitem[Martin et al.(2005)]{martin05}
        Martin N. F., Ibata R. A., Conn B. C., Irwin M. J. \& Lewis G. F. 2005, \pasa\ submitted
%
\bibitem[Mart\'{\i}nez-Delgado et al.(2005)]{martinez-delgado05}
        Mart\'{\i}nez-Delgado D., Butler D. J., Rix H.-W., Franco Y. I. \& Pe\~narrubia J. 2005, \apj\ submitted, {\tt
astro-ph/0410611}
%
\bibitem[McClure-Griffiths et al.(2004)]{mcclure04}
        McClure-Griffiths N., Dickey J., Gaensler B. \& Green A. 2004, \apj\ 607, L127
%
\bibitem[Momany et al.(2004)]{momany04}
        Momany Y., Zaggia S., Bonifacio P., Piotto G., De Angeli F., Bedin L. \& Carraro G. 2004, \aap\ 421, L29
%
\bibitem[Nakanishi \& Sofue(2003)]{nakanishi03}
        Nakanishi H. \& Sofue Y. 2003, PASJ\ 55, 191
%
\bibitem[Newberg et al.(2002)]{newberg02}
        Newberg H. et al. 2002, \apj\ 569, 245
%
\bibitem[Pe\~narrubia et al.(2005)]{penarrubia05}
        Pe\~narrubia J., Mart\'\i{}nez-Delgado D., Rix H.-W., G\'omez-Flechoso M. A., Munn J., Newberg H., Bell E. F.,
Yanny B., Zucker D. \& Grebel E. K. 2005, \apj\ submitted, {\tt astro-ph/0410448}
%
\bibitem[Robin et al.(2003)]{robin03}
       Robin A., Reyl\'e C., Derri\`ere S. \& Picaud S. 2003, \aap\ 409, 523
%
\bibitem[Rocha-Pinto et al.(2004)]{rocha-pinto04}
        Rocha-Pinto H., Majewski S., Skrutskie M., Crane J. \& Patterson R., 2004 \apj\ 615, 732
%
\bibitem[Ruphy et al.(1996)]{ruphy96}
       Ruphy S., Robin A., Epchtein N., Copet E., Bertin E., Fouque P. \& Guglielmo F., 1996 \aap\ 313, L21
%
\bibitem[Schlegel, Finkbeiner \& Davis(1998)]{schlegel98}
        Schlegel D., Finkbeiner D. \& Davis M. 1998, \apj\ 500, 525
%
\bibitem[Soubiran, Bienaym\'e \& Siebert(2003)]{soubiran03}
        Soubiran C., Bienaym\'e O., Siebert A. 2003, A\&A 398, 141
%
\bibitem[Taylor et al.(1996)]{taylor96} 
        Taylor K., Bailey J., Wilkins T., Shortridge K., Glazebrook K., 1996, adass, 5, 195
%
\bibitem[Yanny et al.(2003)]{yanny03}
        Yanny B., Newberg H. et al. 2003, \apj\ 588, 824
%
\bibitem[Zacharias et al.(2004)]{zacharias04}
        Zacharias N., Urban S. E., Zacharias M. I., Wycoff G. L., Hall D. M., Monet D. G. \& Rafferty T. J. 2004, \aj\
127, 3043
%
\end{thebibliography}
\end{document}